\documentclass{article}
\usepackage[utf8]{inputenc}
\usepackage{authblk}
\usepackage{setspace}
\usepackage[margin=1.25in]{geometry}
\usepackage{graphicx}
\graphicspath{ {./figures/} }
\usepackage{subcaption}
\usepackage{amsmath}
\usepackage{lineno}
\usepackage{hyperref}
\usepackage{booktabs}
\usepackage{float}
\usepackage{xcolor}
\usepackage[normalem]{ulem}

\usepackage[round,authoryear]{natbib}
\title{Century-Scale Effect of Climate Change on Meteorite Falls}

\author[1,2*]{Eloy Peña-Asensio}
\author[3,4]{Denis Vida}
\author[5]{Ingrid Cnossen}
\author[2]{Esteban Ferrer}

\affil[1]{Department of Aerospace Science and Technology, Politecnico di Milano, Via La Masa 34, 20156 Milano, Italy.} 
\affil[2]{ETSIAE-UPM-School of Aeronautics, Universidad Politécnica de Madrid, Cardenal Cisneros 3, 28040 Madrid, Spain.}
\affil[3]{Department of Physics and Astronomy, University of Western Ontario, London, Ontario N6A 3K7, Canada.}
\affil[4]{Institute for Earth and Space Exploration, University of Western Ontario, London, Ontario N6A 5B8, Canada}
\affil[5]{British Antarctic Survey, Cambridge, United Kingdom.}
\affil[*]{Address correspondence to: eloy.pena@polimi.it, eloy.peas@gmail.com}

\date{}

\begin{document}

\maketitle

\begin{abstract}
Climate change is inducing a global atmospheric contraction above the tropopause ($\sim$10~km), leading to systematic decrease in neutral air density. The impact of climate change on small meteoroids has already been observed over the last two decades, with documented shifts in their ablation altitudes in the mesosphere ($\sim$50--85~km) and lower thermosphere ($\sim$85--120~km). This study evaluates the potential effect of these changes on meteorite-dropping fireballs, which typically penetrate the stratosphere ($\sim$10--50~km). As a case study, we simulate the atmospheric entry of the fragile Winchcombe carbonaceous chondrite under projected atmospheric conditions for the year 2100 assuming a moderate future emission scenario. Using a semi-empirical fragmentation and ablation model, we compare the meteoroid’s light curve and deceleration under present and future atmospheric density profiles. The results indicate a modest variation of the ablation heights, with the catastrophic fragmentation occurring 300~m lower and the luminous flight terminating 190~m higher. The absolute magnitude peak remains unchanged, but the fireball would appear 0.5 dimmer above $\sim$120~km. The surviving meteorite mass is reduced by only 0.1 g. Our findings indicate that century-scale variations in atmospheric density caused by climate change moderately influence bright fireballs and have a minimal impact on meteorite survival.
\end{abstract}


\section{Introduction}

The increase in atmospheric CO$^{\text{2}}$ concentration causes net warming in the troposphere ($\sim$0–10~km), but results in cooling from the stratosphere ($\sim$10–50~km) upward \citep{Manabe1967, Manabe1975, Roble1989}. This has widespread effects on the climate of the stratosphere ($\sim$10–50~km), mesosphere ($\sim$50–85~km), and thermosphere ($\sim$85–500~km), including thermal contraction: cooling of the middle and upper atmosphere causes this part of the atmosphere to shrink, so that the atmospheric mass density at high altitudes shows a multi-decadal decrease \citep[e.g.,][]{Emmert2015, Weng2020, Cnossen2024}. This has important practical consequences for the lifetime of space debris in low Earth orbit and long-term satellite mission planning \citep[e.g.,][]{Brown2021, Brown2024}.

At lower altitudes, between the mesosphere and the lower thermosphere, atmospheric contraction has also significantly affected meteor ablation altitudes---the height at which a small space rock (meteoroid) enters the atmosphere and produces visible radiation due to its interaction with the air \citep{Koschny2017JIMO4591K}. A thinning upper atmosphere should, in principle, lead to a systematic lowering of meteor ablation height, as the altitude at which meteoroids ablate depends on atmospheric mass density. 

Observational studies using radars have consistently documented such trends in small impacting meteoroids (magnitude +8 and fainter; roughly 100–2000~$\mu$m in diameter), providing empirical evidence that the shrinking atmosphere is altering the height distribution of meteors detected by ground-based systems. Early observations by \citet{Clemesha2006JASTP681934C} revealed fluctuations in meteor centroid altitudes in Cachoeira Paulista, Brazil (23°S, 45°W), with reported decreases ranging from 300 to 800~m per decade between 2000 and 2005. \citet{Jacobi2014AdRS12161J} examined meteor altitudes at Collm, Germany (51.3°N, 13.0°E) and reported a trend of -560~m per decade, with an additional solar cycle effect of +450~m per 100 solar flux units, indicating that both cooling and solar activity modulate the observed altitudes. Similarly, \citet{Lima2015JASTP133139L} observed a comparable decrease of 380~m per decade in Cachoeira Paulista, Brazil, after accounting for solar variability. More recently, a data set from Tirupati, India (13.63°N, 79.4°E) showed a decrease of 228~m per decade in meteor peak altitude over an 11-year period \citep{VenkatRatnam2024}. 

The sensitivity of meteor ablation to atmospheric density fluctuations is further demonstrated by the response of meteor peak flux altitudes to planetary wave activity. \citet{Stober2012JASTP7455S} showed that large-scale atmospheric perturbations can cause measurable shifts in meteor heights from four months of observations made in Collm, Juliusruh, Germany (54.6°N, 13.4°E), and Andenes, Norway (69.3°N, 16.0°E). Extending these findings to longer timescales, \citet{Stober2014GeoRL416919S} used a decade of meteor radar observations from the Canadian Meteor Orbit Radar (CMOR) in Ontario, Canada (43.3° N, 80.8° W), to estimate changes in neutral air density and reported a decrease of 5.8\% per decade at approximately 91~km altitude, consistent with expectations from global climate models.

Recent large-scale analyses provide further confirmation that this trend is not limited to specific locations. Using over 20 years of data from 12 meteor radars distributed over a broad range of latitudes, \citet{Dawkins2023} reported that meteor ablation altitudes are decreasing at all sites, with trends ranging from 10 to 818~m per decade. This study also found a strong correlation between meteor heights and solar activity at most stations, but confirmed that multi-decadal mesospheric cooling due to increasing greenhouse gas concentrations is systematically lowering the altitude at which meteors are detected.

Although these studies primarily focus on small meteoroids detected by radar, similar mechanisms may apply to larger objects, including meteorite-dropping bright meteors (or fireballs), as the atmospheric density profile can have significant effects on meteorite survival dynamical modeling \citep{LyytinenGritsevich2016}. A reduction in atmospheric density could alter the depth at which fireballs begin and end their luminous phase, potentially affecting the likelihood of meteorite survival. It is therefore conceivable that climate change in the entire atmosphere could affect the meteorite fall rate as well, especially of fragile carbonaceous chondrites which require specific conditions to survive atmospheric flight \citep{brown2002entry}. The purpose of this study is to investigate this further, using the Winchcombe carbonaceous chondrite meteorite \citep{Gattacceca2022MPS572102G} fall as a case study.

The remainder of the paper is organized as follows. We first detail the methodology in Section \ref{sec:methods} to then present and discuss the results in Section \ref{sec:results}. Finally, conclusions can be found in Section \ref{sec:conclusions}.

\section{Methods} \label{sec:methods}

This section outlines the methodology used to assess the potential effects of atmospheric density variations on meteorite-dropping fireballs, using the Winchcombe meteorite fall as a case study. First, we describe the observational data and key physical characteristics of the Winchcombe meteorite fall (\S\ref{sec:winchcombe}), which serve as the foundation for the subsequent modeling. Next, we detail the fragmentation and erosion model applied to reproduce the observed meteoroid’s atmospheric entry (\S\ref{sec:atm_entry}). Then we present the methodology used to estimate century-scale trends in atmospheric density based on climate change simulations (\S\ref{sec:clim_change_pred}). Finally, we use these predictions to simulate the Winchcombe fireball under different atmospheric conditions, allowing us to evaluate possible modifications in its luminous trajectory and meteorite survival (\S\ref{sec:winchcombe_clim_change}).

\subsection{The Winchcombe meteorite} \label{sec:winchcombe}

The Winchcombe meteorite fall on February 28, 2021, in the United Kingdom, represents one of the best-documented cases of a fireball with recovered meteorites \citep{King2022SciA8Q3925K, McMullan2024MPS59927M, Russell2024MPS59973R, Suttle2024MPS591043S}. The fireball was recorded by 16 optical stations, including dedicated meteor camera networks and casual video recordings, providing accurate measurements of atmospheric trajectory, fireball dynamics, light curve, and fragmentation. The event was observed by multiple fireball networks, including the UK Fireball Network (UKFN) within the Global Fireball Observatory \citep[GFO,][]{Devillepoix2020PSS19105036D}, the System for Capture of Asteroid and Meteorite Paths (SCAMP) within the French Fireball Recovery and InterPlanetary Observation Network \citep[FRIPON,][]{Colas2020AA644A53C}, the Global Meteor Network \citep[GMN,][]{Vida2021MNRAS5065046V}, the UK Meteor Observation Network \citep[UKMON,][]{CampbellBurns2014JIMO42139C}, the Network for Meteor Triangulation and Orbit Determination \citep[NEMETODE,][]{Stewart2013JIMO4184S}, and the AllSky7 network \citep{Hankey2020pimoconf21H}, with the UK Fireball Alliance \citep[UKFAll,][]{Daly2020EPSC14705D} coordinating the efforts in the UK.

The progenitor meteoroid had a very small estimated initial mass of approximately 13~kg and entered Earth's atmosphere with a relatively low velocity of 13.9~km/s. The fireball was first detected at an altitude of 90.6~km and remained observable down to 27.6~km, where its velocity dropped to approximately 3~km/s \citep{McMullan2024MPS59927M}. The meteoroid experienced a major fragmentation event at 35~km altitude, producing multiple trackable fragments. 

Winchcombe is classified as a CM2 carbonaceous chondrite, a rare and fragile meteorite type with a high volatile and water content. The meteorite has a bulk density of approximately 2090~kg/m\textsuperscript{3}, consistent with micro-X-ray computed tomography measurements of recovered fragments \citep{King2022SciA8Q3925K}. The total recovered mass from the strewn field was approximately 0.6~kg, with the largest fragment weighing 0.319~kg.


Winchcombe is an ideal case study for atmospheric effects on meteoroid survival, not only due to its high-quality observational data but also its fragility. While C-type asteroids comprise 13–20\% of near-Earth objects \citep{Binzel2019Icar32441B, Ieva2020AA644A23I, Hromakina2021AA656A89H}, carbonaceous chondrites make up only $\sim$4\% of recovered meteorites\footnote{\url{https://www.lpi.usra.edu/meteor/}}, indicating most disintegrate during entry. Dynamical modeling suggests that carbonaceous materials may constitute over 50\% of the terrestrial impactor population \citep{Broz2024AA}, further reinforcing this survival bias. Their extreme weakness---some, like Tagish Lake, crumble by hand \citep{Brown2000}---makes them particularly susceptible to both thermal fragmentation in space and atmospheric filtering during entry \citep{Shober2025}. If climate change influences meteoroid fragmentation and ablation, carbonaceous chondrites would be the first to reveal these effects.


\subsection{Meteoroid atmospheric entry model} \label{sec:atm_entry}

The atmospheric entry of the meteoroid was analyzed using a semi-empirical ablation model developed by \citet{Borovicka2013MPS481757B}, following the manual modeling procedure detailed in \citet{Borovicka2020AJ16042B} and using the implementation by \cite{Vida2023NatAs7318V}.

The initial meteoroid is modeled as a classical single body that fragments at manually determined points. The physical parameters of the initial meteoroid and individual fragments are also manually estimated. The fragmentation points are informed by brightness increases in the light curve (i.e., flares) and a consequent increase in the observed deceleration. Most fragmentations were modeled as the release of an eroding fragment (see \citet{Borovicka2015astebook257B} for more details), which is a body that rapidly erodes by the release of mm-sized grains. The grain masses are distributed according to a power law (a differential mass index of $s = 2.0$ is assumed) within a given range of masses. The amount of erosion is regulated by the erosion coefficient $\eta$, which determines how much mass is eroded from the fragment per unit of kinetic energy loss.
Both the main body and fragmented grains are modeled using the classical single-body ablation equations \citep{Ceplecha1998SSRv84327C}:

\begin{equation}
    \frac{dv}{dt} = -K m^{-1/3} \rho_{\text{air}} v^2 \,,
\end{equation}
\begin{equation}
    \frac{dm_a}{dt} = -K \sigma m^{2/3} \rho_{\text{air}} v^3 \,,
\end{equation}

\noindent where $K$ is the shape density coefficient, $m$ is the meteoroid mass ($m_a$ is the ablated mass), $v$ is the velocity, and $\rho_{\text{air}}$ is the atmospheric density taken from the NRLMSISE-00 model \citep{Picone2002JGRA1071468P}. The ablation coefficient, $\sigma$, represents the fraction of the kinetic energy converted into mass loss. As the density and shape of the meteoroid cannot be measured independently, the coefficient $K$ is defined as:

\begin{equation}
    K = \frac{\Gamma A}{\rho_m^{2/3}} \,,
\end{equation}

\noindent where $\Gamma$ is the drag coefficient (typically set at 1), $A$ is the shape factor (taken as 1.21 for spheres), and $\rho_m$ is the bulk density of the meteoroid or individual grains.

The equations of motion and ablation were numerically integrated using a fourth-order Runge-Kutta scheme with a time step of 2 ms. The integration of individual fragments was stopped if the mass of the fragment fell below $10^{-14}$~kg or if the velocity dropped below 2.5~km/s, the assumed ablation limit \citep{Ceplecha1998SSRv84327C}.

The total luminosity generated by ablation is computed as:

\begin{equation}
    I = -\tau \frac{v^2}{2} \frac{dm_a}{dt} + m v \frac{dv}{dt} \,,
\end{equation}

\noindent where $\tau$ is the luminous efficiency. The luminous efficiency function used in this work followed the formulation given by \citet{Borovicka2020AJ16042B}.

For cases where the body or its fragments undergo erosion, the mass loss due to erosion follows a formulation similar to that of ablation:

\begin{equation}
    \frac{dm_e}{dt} = -K \eta m^{2/3} \rho_{\text{air}} v^3 \,,
\end{equation}

\noindent where $\eta$ is the erosion coefficient. The eroded mass is distributed into smaller fragments whose masses follow a power law distribution defined by the mass index $s$ and the range of ejected masses. These fragments ablate independently and single-body equations are integrated for each, with the total fireball luminosity being the sum of luminosities of all ablating fragments at a given time.

The total mass loss for an object in the simulation at any given time is the sum of ablation and erosion contributions:

\begin{equation}
    \frac{dm}{dt} = \frac{dm_a}{dt} + \frac{dm_e}{dt} \,.
\end{equation}

For the bulk density of the grains produced by erosion, a value of $3000$~kg/m$^3$ is used, suitable for refractory silicate materials. A complete description of the method implemented for grain mass distribution can be found in the section Methods in \citet{Vida2023NatAs7318V}.\footnote{The corresponding code implementation is openly available at \url{https://github.com/wmpg/WesternMeteorPyLib}.}

\subsection{Atmospheric density projection} \label{sec:clim_change_pred}

We used a simulation by \citet{Cnossen2022} with the Whole Atmosphere Community Climate Model – eXtended (WACCM-X) for the period 2015-2070 to estimate the future century-scale trend in the global mean mass density of air. This simulation followed the shared socio-economic pathway 2-4.5 \citep{ONeill2016}, a “middle of the road” scenario, and also included realistic solar and geomagnetic activity variations, based on \citet{Matthes2017}. In the thermosphere, the approximately 11-year solar cycle produces large variations in density, which far exceed the magnitude of the multi-decadal trend. To extract the century-scale trend in the global mean mass density, the model data was fitted to the following linear regression model:

\begin{equation}
\rho^{\prime}_{\text{air}} = a + b \cdot F10.7a + c \cdot (F10.7a)^2 + d \cdot K_p + e \cdot year,
\label{linreg}
\end{equation}

\noindent where $\rho^{\prime}_{\text{air}}$ is the global mean mass density at a given altitude, $F10.7a$ is the 81-day average of the $F10.7$ index of solar activity, $K_p$ is the geomagnetic activity index, and $a$, $b$, $c$, $d$, and $e$ are coefficients to fit, with $e$ corresponding to the trend. 
 
While altitudes below the thermosphere are progressively less affected by solar and geomagnetic activity variations, the same formulation was used here, as the results for lower altitudes were not sensitive to omission/inclusion of the solar and geomagnetic activity terms. However, results in the thermosphere do vary somewhat depending on the exact time interval used, which is likely due to incomplete removal of solar activity effects. We therefore followed the approach of \citet{Cnossen2020, Cnossen2022} to use the average of the trends obtained for the 11 periods starting in January 2015 and ending in December 2060-December 2070 (with each subsequent period being 12 months longer than the last). The global mean trend was used here, as mass density trends do not tend to vary much with location \citep[e.g.,][]{Cnossen2020}. 

\subsection{Simulating Winchcombe fall with another atmosphere density profile} \label{sec:winchcombe_clim_change}

The semi-empirical model described in Section \ref{sec:atm_entry} was successfully applied to reconstruct the fragmentation behavior and physical properties of many meteorite-dropping fireballs \citep{Borovicka2015MPS501244B, Borovicka2017PSS143147B, Borovicka2019MPS541024B, Borovicka2020AJ16042B, Brown2023MPS581773B, McMullan2024MPS59927M}. This body of works shows that the fragmentation of chondritic meteorite-dropping meteoroids is primarily driven by mechanical stress, particularly due to aerodynamic loading as the meteoroid passes through the atmosphere. Fragmentation typically occurs when the dynamic pressure exerted by the atmosphere exceeds the cohesive strength of the meteoroid material. The dynamic pressure ($P_{\text{dyn}}$) can be expressed as:

\begin{equation}
    P_{\text{dyn}} = \Gamma v^2 \rho_{air}.
\end{equation}

The Winchcombe fireball experienced an exceptionally low peak atmospheric dynamic pressure of approximately $\sim$0.6~MPa, a result of its favorable entry conditions, allowing it to survive the atmospheric flight. Its structural fragility became evident when it underwent near-catastrophic fragmentation at a significantly lower pressure of only $\sim$0.07~MPa.

To model the expected behavior of the Winchcombe meteoroid in a different atmosphere, we use the same initial conditions and physical properties while allowing the model to integrate the atmospheric flight, computing new light and deceleration curves. We enforce identical fragmentation events (type and mass loss) at the same dynamic pressure. This is achieved by mapping the heights between different atmospheric density profiles.

In the original study of the Winchcombe fireball, a 7$^{\text{th}}$ degree polynomial was fitted to the logarithm of the atmospheric density profile to speed up the sampling of the atmospheric model at every integration step. In this analysis, we refine the approach employing an 11$^{\text{th}}$ degree polynomial to capture finer details. However, the differences between these approximations remain minimal.

To estimate a density profile for 2100, we apply the predicted atmospheric density trend to the year 2100 (\S\ref{sec:clim_change_pred}) in the average trajectory location of the Winchcombe fireball (51ºN, 2.6ºW) at the same hour, day, and month. Throughout this study, we refer to \textit{2021} as the observed event (the reference one), using the recorded atmospheric conditions, and \textit{2100} as the projected scenario based on the projected climate change. In this study, \textit{2021} denotes the reference year with recorded atmospheric conditions on the event day, and \textit{2100} denotes the projected climate‐change scenario. All differences between model results are defined as projected (2100) minus reference (2021).

\section{Results} \label{sec:results}

\begin{figure}[H]
    \centering
    \includegraphics[width=0.85\textwidth]{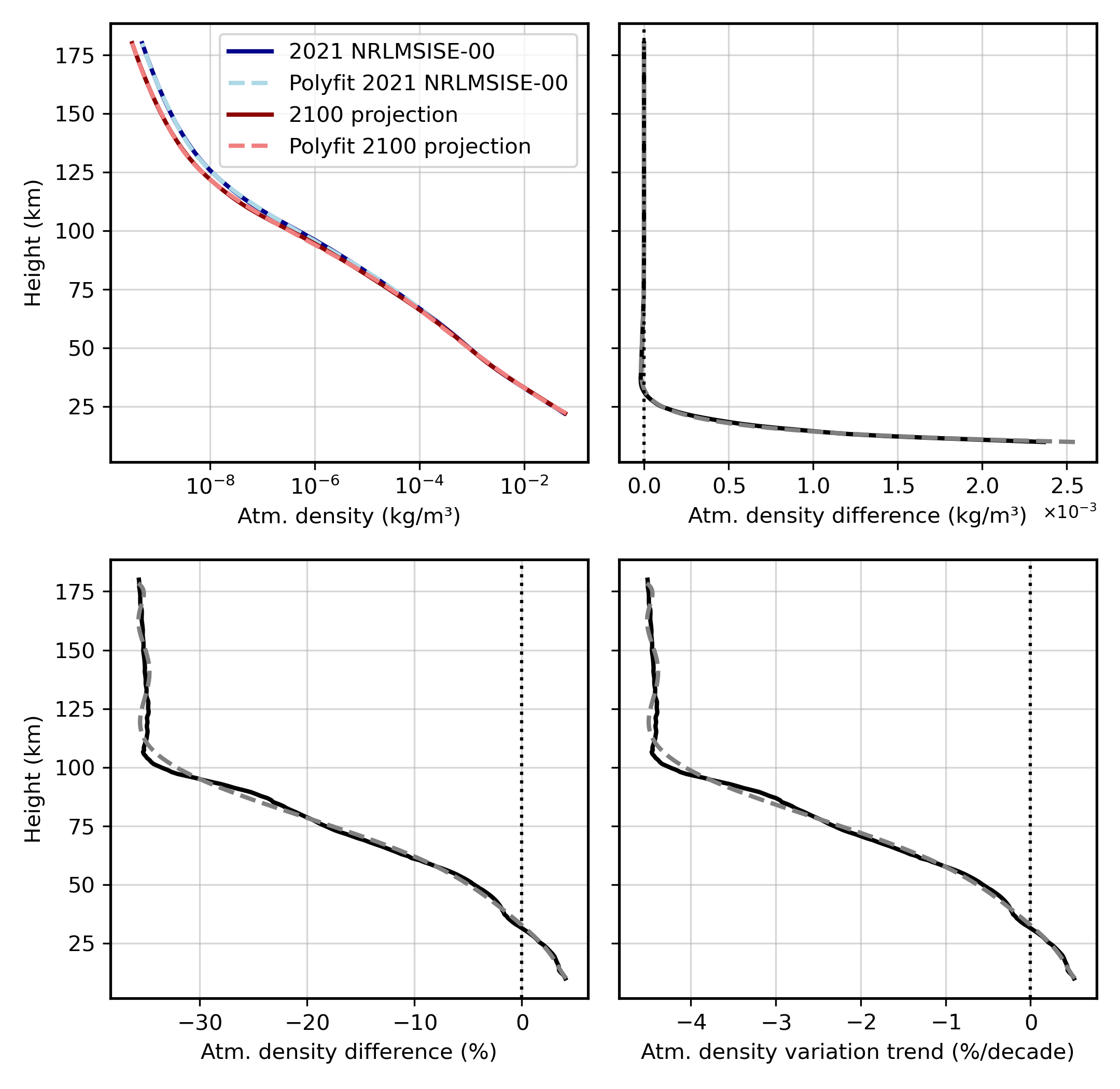}
    \caption{Top left: atmospheric density for 2021 (blue) and 2100 projection (red), with 11th-degree polynomial fits (lighter dashed). Top right: absolute difference (solid black) and its fit (dashed). Bottom left: percent difference (solid) and fit (dashed). Bottom right: trend in \% per decade (solid) and fit (dashed). Differences are computed as projected (2100) minus reference (2021).
    }
    \label{fig:atmospheres}
\end{figure}

Figure~\ref{fig:atmospheres} shows two profiles: the reference atmospheric density profile sampled from the NRLMSISE-00 model appropriate for the time and location of the Winchcombe fireball (2021) and the projected profile for the year 2100, along with their respective polynomial fits. The polynomial approximation shows increasing relative residuals above 75~km. While this represents a larger relative deviation compared to lower altitudes, the absolute impact on fragmentation and deceleration is minimal at such heights due to the low densities involved and the fact that luminosity is primarily generated by drag. Although the effects of climate change on atmospheric density are indeed more pronounced above 100~km due to the cumulative effect of atmospheric contraction below---projected to increase by $\sim35\%$ by 2100---this occurs at altitudes where atmospheric densities are low enough that the polynomial model retains sufficient accuracy to assess the entry behavior of meteoroids.

In addition, a trend reversal is observed, with the atmospheric density increasing below approximately 37~km. However, as shown in the top right panel of Figure~\ref{fig:atmospheres}, absolute differences in density are only noticeable below 25~km, and even at these lower altitudes, the variations remain relatively small ($\sim5\%$) and on par with diurnal and seasonal variations of up to $25\%$ \citep{vida2021high}.

\begin{figure}[H]
    \centering
    \includegraphics[width=0.85\textwidth]{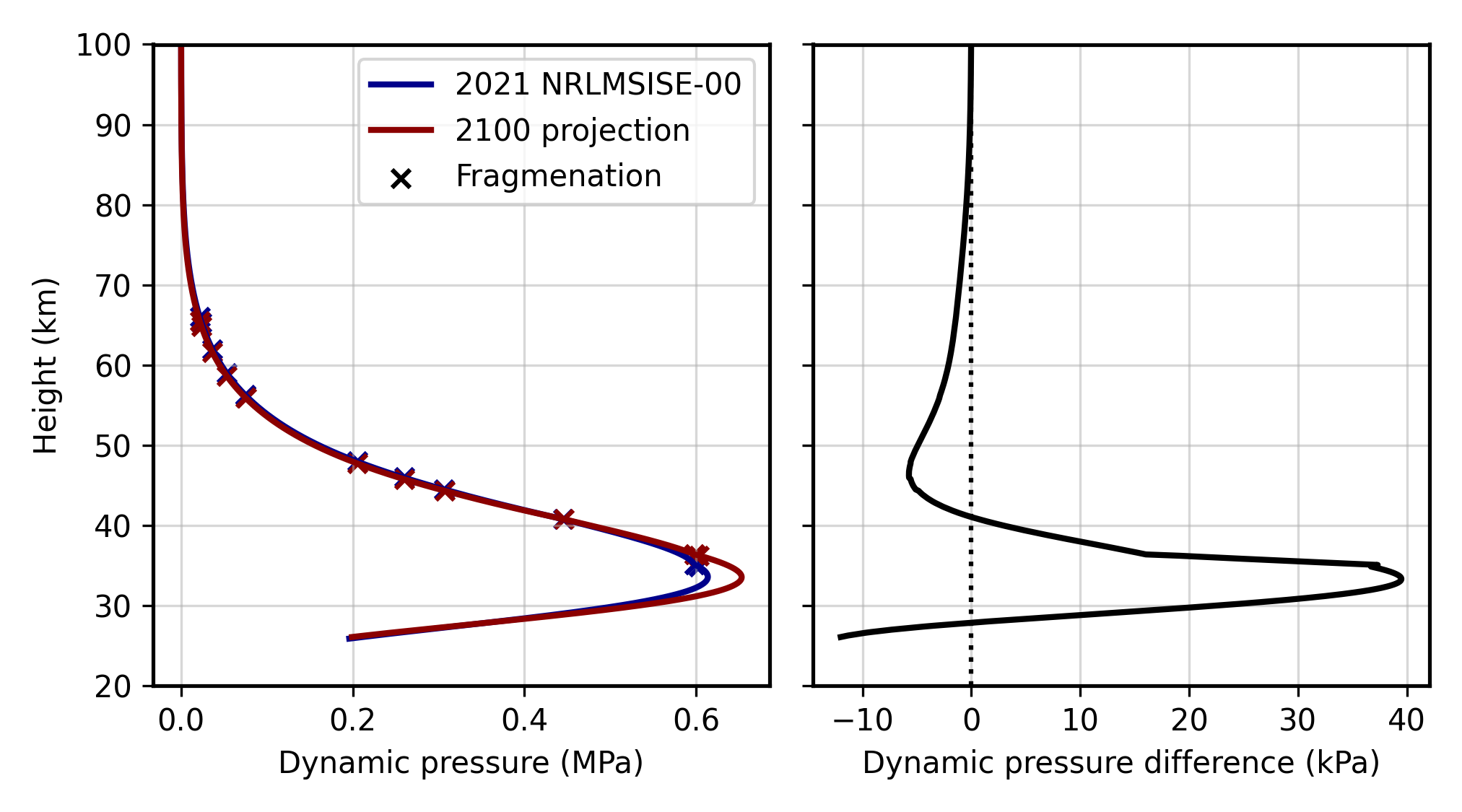}
    \caption{Dynamic pressure as a function of atmospheric height for the Winchcombe fireball (blue lines) and the simulated Winchcombe fall for the year 2100 considering climate change (red lines). The eleven fragmentation events are marked with crosses. The right panel represents the differences in dynamic pressure (black line). Difference is computed as projected (2100) minus reference (2021). 
}
    \label{fig:height_dynmass_2}
\end{figure}

Figure~\ref{fig:height_dynmass_2} presents the dynamic pressure as a function of atmospheric height for the Winchcombe fireball (2021) and the simulated scenario for 2100 under climate change conditions. The right panel highlights the differences in dynamic pressure between the two scenarios. Above 41~km, the meteoroid experiences a lower dynamic pressure, with differences reaching up to 5~kPa. However, below 37~km, the dynamic pressure increases in the 2100 scenario and peaks at a maximum difference of 40~kPa (0.614 vs 0.653~MPa). Toward the final moments of the luminous flight, the dynamic pressure decreases again and remains 12~kPa lower in 2100 compared to 2021. This behavior reflects the interplay between atmospheric density and deceleration. In the left panel of Figure~\ref{fig:height_dynmass_2}, fragmentation events remain well aligned in terms of dynamic pressure, but occur at different altitudes to ensure consistency (see \S\ref{sec:winchcombe_clim_change}).

\begin{figure}[H]
    \centering
    \includegraphics[width=0.85\textwidth]{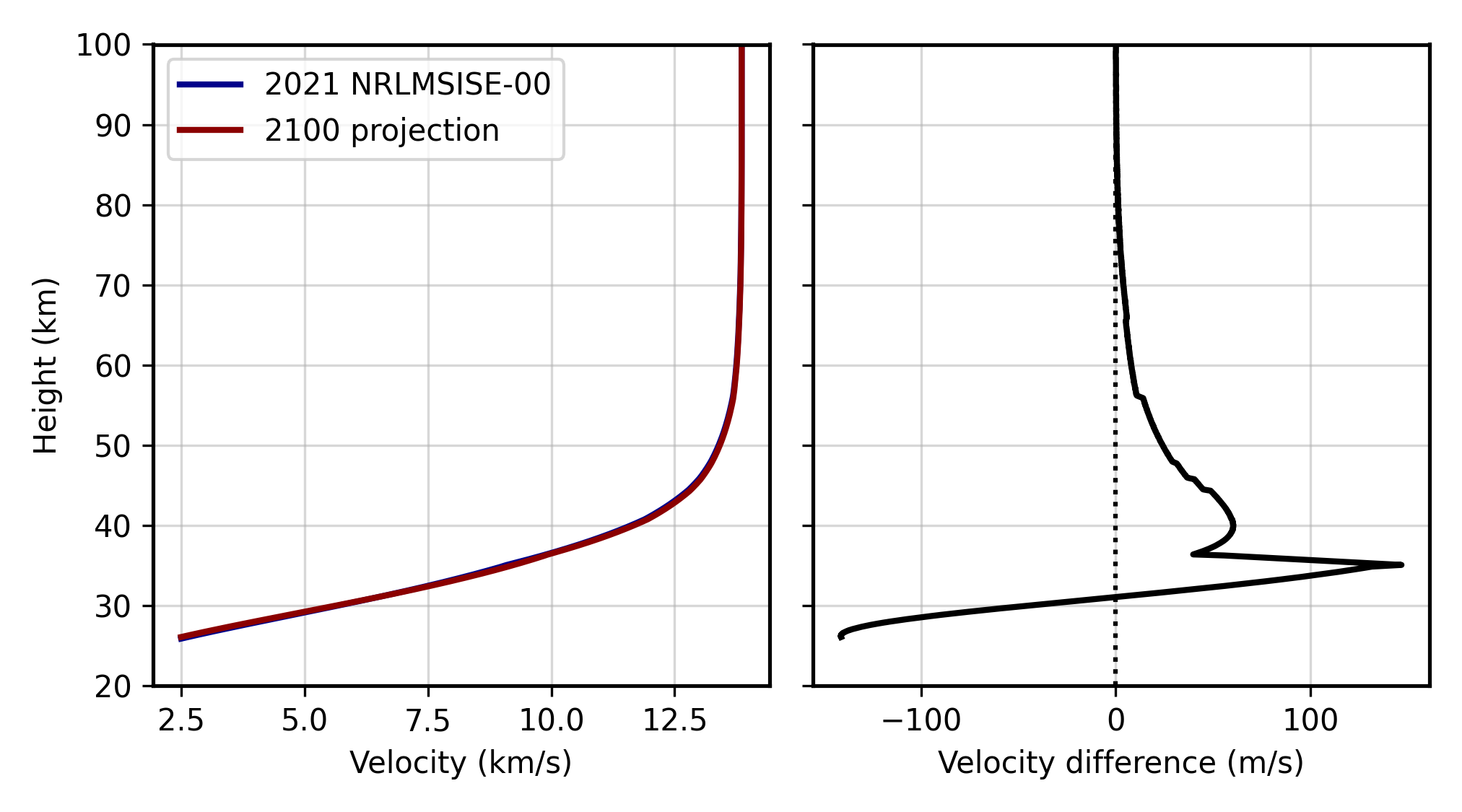}
    \caption{Velocity of the main fragment as a function of height for the Winchcombe fireball (blue line) and the simulated Winchcombe fall for the year 2100 considering climate change (red line). The right panel shows the velocity difference (black line). Difference is computed as projected (2100) minus reference (2021).
}
    \label{fig:velocity}
\end{figure}

\begin{figure}[H]
    \centering
    \includegraphics[width=0.85\textwidth]{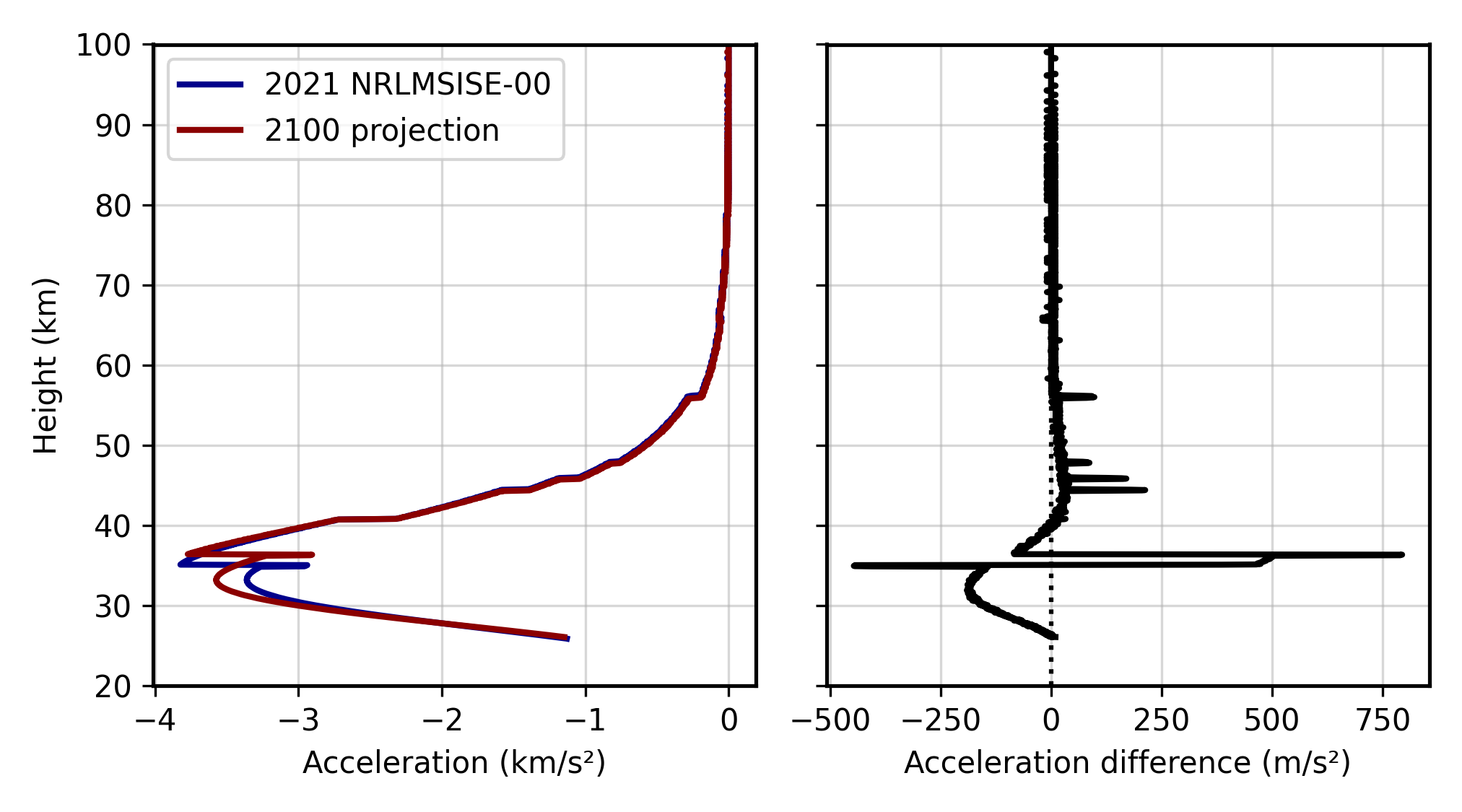}
    \caption{Acceleration of the main fragment as a function of height for the Winchcombe fireball (blue line) and the simulated Winchcombe fall for the year 2100 considering climate change (red line). The right panel shows the acceleration difference (black line). Difference is computed as projected (2100) minus reference (2021).
}
    \label{fig:deceleration}
\end{figure}

Figure~\ref{fig:velocity} shows the velocity of the main fragment as a function of height for the Winchcombe fireball (2021) and the simulated scenario for 2100. The velocity remains consistently higher throughout most of the trajectory in the 2100 scenario. This behavior is expected, as the atmosphere in 2100 is generally less dense, allowing the meteoroid to experience lower deceleration at higher altitudes. However, once it reaches approximately 37~km, the increased atmospheric density in the 2100 scenario leads to a more rapid deceleration, as seen in Figure~\ref{fig:deceleration}, reducing the velocity difference in the final phase of the luminous flight. The velocity difference varies between -147 and 141~m/s when comparing both scenarios.

\begin{figure}[H]
    \centering
    \includegraphics[width=0.85\textwidth]{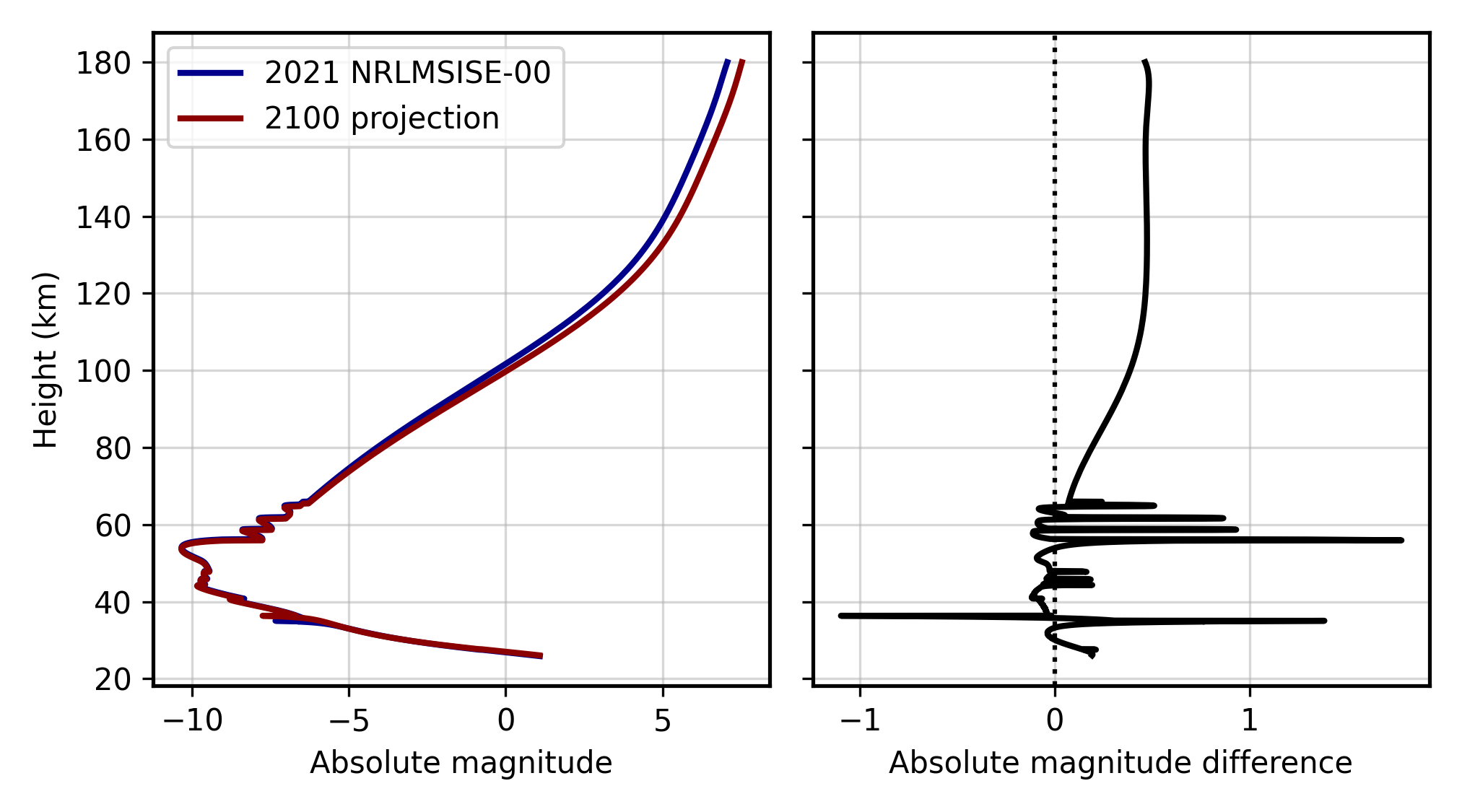}
    \caption{Absolute magnitude (brightness) of the main fragment as function of height for the Winchcombe fireball (blue line) and the simulated Winchcombe fall for the year 2100 considering climate change (red line). The right panel shows the absolute magnitude difference (black line). Difference is computed as projected (2100) minus reference (2021).
}
    \label{fig:magnitude}
\end{figure}

Figure~\ref{fig:magnitude} presents the light curve of the main fragment as a function of height for the Winchcombe fireball (2021) and the simulated scenario for 2100 under climate change conditions. The absolute magnitude of a fireball is a normalized brightness at a fixed range of 100~km and without the effects of atmospheric extinction, with the brightness of the star Vega used as the zero point. Lower values indicate greater brightness, and a difference of one magnitude is equivalent to a brightness difference of $2.512\times$. 

A clear difference is observed above 120~km, where the 2100 simulation is consistently fainter by approximately 0.5 magnitudes. This difference decreases at lower altitudes, with only brief peaks in brightness variation. These peaks are attributed to the different altitudes at which fragmentation events occur in each case. Despite these variations, the maximum brightness reached in both scenarios remains virtually the same, just brighter than magnitude -10. For example, a CCD camera with a limiting absolute magnitude of $+3$ would first detect the Winchcombe fireball 3~km lower in altitude in 2100.

\begin{figure}[H]
    \centering
    \includegraphics[width=0.6\textwidth]{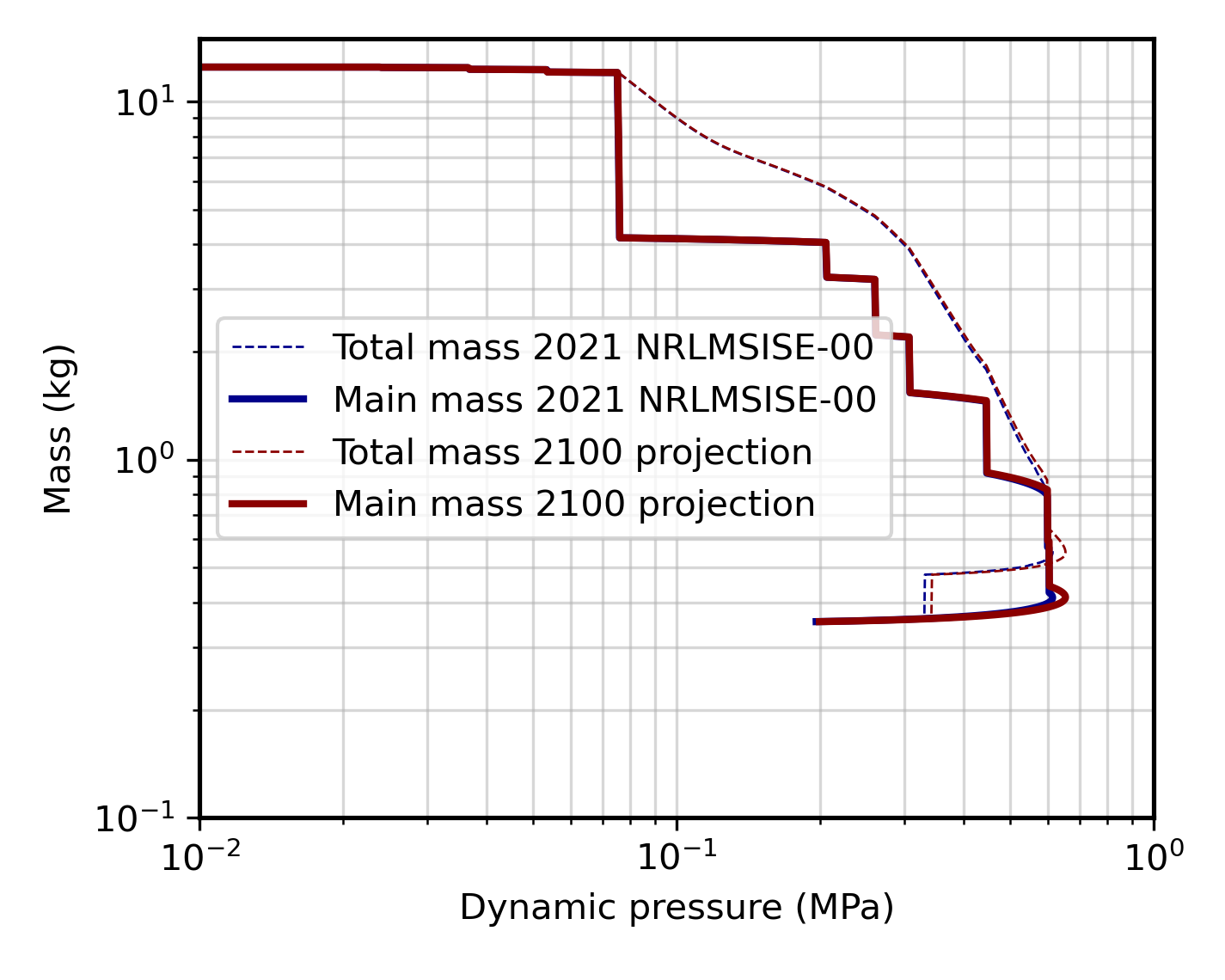}
    \caption{Mass of the main fragment (solid line) and the total mass (dashed line) versus the dynamic pressure for the Winchcombe fireball (blue lines) and the simulated Winchcombe fall for the year 2100 considering climate change (red lines).
}
    \label{fig:height_dynmass_1_log}
\end{figure}

Beginning with an initial mass of 12.5~kg, the meteoroid underwent only minor fragmentation above 60~km in altitude in both scenarios, at dynamic pressures not exceeding 0.06~MPa. For a complete representation of the atmospheric fragmentation history, Figure~\ref{fig:height_dynmass_1_log} shows the evolution of the main fragment mass (solid line) and the total mass (dashed line) as a function of dynamic pressure. The blue lines correspond to the Winchcombe fireball (2021), while the red lines depict the simulated fall for the year 2100 under climate change conditions.

In the 2100 simulation, the first fragmentation occurs 820~m lower than in the 2021 case, and the catastrophic fragmentation takes place 300~m lower. In contrast, the final luminous flight altitude is 190~m higher, and the meteorite mass is reduced by 0.13 g (-0.037\%) compared to the 2021 scenario. In Table~\ref{tab:fragmentation}, we compile the fragmentation behavior modeled for the Winchcombe fireball (2021) and the simulated fall under the 2100 climate change scenario, including the final height and the surviving meteorite mass at the end of the ablation. It is reasonable to assume that for larger and stronger meteoroids, the impact of atmospheric density variations on meteorite survival would be even smaller.

\begin{table}[H]
\caption{Modeled fragmentation behavior for the Winchcombe fireball (2021) and the simulated Winchcombe fall considering climate change (2100). The last row represent the final height and mass of the main fragment at the end of ablation. Abbreviations: dust ejection (D); eroding fragment (EF); single-body fragment (F).}
\label{tab:fragmentation}
\begin{tabular}{cc cc cc c cc cc} 
\hline
\multicolumn{2}{c}{Height (km)} & \multicolumn{2}{c}{Velocity (km/s)} & \multicolumn{2}{c}{$P_{dyn}$ (MPa)} & Frag. & \multicolumn{2}{c}{Mass loss (kg)} & \multicolumn{2}{c}{Meteorite (g)} \\
2021 & 2100 & 2021 & 2100 & 2021 & 2100 &  & 2021 & 2100 & 2021 & 2100 \\
\midrule
65.30 & 64.85 & 13.80 & 13.80 & 0.024 & 0.024 & EF & 0.025 & 0.025 & $--$ & $--$ \\
62.00 & 61.60 & 13.77 & 13.78 & 0.037 & 0.037 & EF & 0.124 & 0.124 & $--$ & $--$ \\
59.00 & 58.66 & 13.74 & 13.74 & 0.053 & 0.053 & EF & 0.184 & 0.184 & $--$ & $--$ \\
56.25 & 55.94 & 13.69 & 13.69 & 0.075 & 0.075 & EF & 4.343 & 4.343 & $--$ & $--$ \\
56.20 & 55.91 & 13.68 & 13.69 & 0.075 & 0.076 & EF & 3.551 & 3.552 & $--$ & $--$ \\
48.00 & 47.76 & 13.23 & 13.24 & 0.205 & 0.206 & EF & 0.809 & 0.809 & $--$ & $--$ \\
46.00 & 45.78 & 13.02 & 13.03 & 0.260 & 0.260 & EF & 0.957 & 0.957 & $--$ & $--$ \\
44.50 & 44.36 & 12.79 & 12.81 & 0.307 & 0.307 & EF & 0.660 & 0.661 & $--$ & $--$ \\
40.80 & 40.80 & 11.90 & 11.96 & 0.445 & 0.446 & EF & 0.539 & 0.541 & 1.12 & 1.05 \\
35.10 & 36.40 & 9.10 & 9.93 & 0.598 & 0.599 & D & 0.222 & 0.231 & $--$ & $--$ \\
34.90 & 36.26 & 9.00 & 9.88 & 0.603 & 0.604 & 3$\times$F & 0.142 & 0.148 & 3$\times$117.67 & 3$\times$117.65 \\
25.89 & 26.08 & $--$ & $--$ & $--$  & $--$  & $--$       & $--$  & $--$   & 354.12  & 353.99  \\
\bottomrule
\end{tabular}
\end{table}

\section{Conclusions}\label{sec:conclusions}

We have investigated the potential impact of climate change on meteorite falls by modeling the atmospheric entry of the Winchcombe meteoroid under predicted conditions for the year 2100 assuming a moderate future emission scenario. The combination of high-quality observational data, including light and deceleration curves, together with detailed fragmentation modeling and the inherent fragility of the meteoroid, makes the Winchcombe event an ideal case study to evaluate the impact of atmospheric density variations on larger meteoroids that produce meteorites.

The analysis indicates that, while the projected 2100 atmosphere exhibits a systematic reduction in density at high altitudes, the overall atmospheric flight is only modestly affected. Under these conditions, a CCD camera with a limiting absolute magnitude of $+3$ would detect the Winchcombe fireball 3~km lower in altitude. The first fragmentation occurs 820~m lower, and the catastrophic fragmentation takes place 300~m lower, while the luminous flight ends 190~m higher. Furthermore, the surviving meteorite mass is reduced by approximately 0.1 g compared to the real case.

In contrast to small meteoroids, which completely ablate at higher altitudes and are therefore more strongly affected by changes in atmospheric density, larger and slower meteoroids do not experience significant alterations in their ablation process. Overall, these results suggest that century-scale changes in atmospheric density due to climate change have a moderate impact on bright fireballs and a limited effect on meteorite survival.

\section*{Acknowledgments}
EP-A acknowledges financial support from the LUMIO project funded by the Agenzia Spaziale Italiana (2024-6-HH.0). DV was supported in part by the NASA Meteoroid Environment Office under cooperative agreement 80NSSC24M0060. IC was supported by a Natural Environment Research Council (NERC) Independent Research Fellowship (NE/R015651/1). EF acknowledges the funding received by the Grant DeepCFD (Project No. PID2022-137899OB-I00) funded by MICIU/AEI/10.13039/501100011033 and by ERDF, EU.

\bibliography{0Bib_Eloy.bib}

\begin{thebibliography}{}

\bibitem[{Binzel} et~al., 2019]{Binzel2019Icar32441B}
{Binzel}, R.~P., {DeMeo}, F.~E., {Turtelboom}, E.~V., {Bus}, S.~J., {Tokunaga}, A., {Burbine}, T.~H., {Lantz}, C., {Polishook}, D., {Carry}, B., {Morbidelli}, A., {Birlan}, M., {Vernazza}, P., {Burt}, B.~J., {Moskovitz}, N., {Slivan}, S.~M., {Thomas}, C.~A., {Rivkin}, A.~S., {Hicks}, M.~D., {Dunn}, T., {Reddy}, V., {Sanchez}, J.~A., {Granvik}, M., and {Kohout}, T. (2019).
\newblock {Compositional distributions and evolutionary processes for the near-Earth object population: Results from the MIT-Hawaii Near-Earth Object Spectroscopic Survey (MITHNEOS)}.
\newblock {\em Icarus}, 324:41--76.

\bibitem[{Borovi{\v{c}}ka} et~al., 2019]{Borovicka2019MPS541024B}
{Borovi{\v{c}}ka}, J., {Popova}, O., and {Spurn{\'y}}, P. (2019).
\newblock {The Maribo CM2 meteorite fall{\textemdash}Survival of weak material at high entry speed}.
\newblock {\em Meteoritics \& Planetary Science}, 54(5):1024--1041.

\bibitem[{Borovi{\v{c}}ka} et~al., 2015a]{Borovicka2015astebook257B}
{Borovi{\v{c}}ka}, J., {Spurn{\'y}}, P., and {Brown}, P. (2015a).
\newblock {Small Near-Earth Asteroids as a Source of Meteorites}.
\newblock In {Michel}, P., {DeMeo}, F.~E., and {Bottke}, W.~F., editors, {\em Asteroids IV}, pages 257--280. The University of Arizona Press.

\bibitem[{Borovi{\v{c}}ka} et~al., 2017]{Borovicka2017PSS143147B}
{Borovi{\v{c}}ka}, J., {Spurn{\'y}}, P., {Grigore}, V.~I., and {Svore{\v{n}}}, J. (2017).
\newblock {The January 7, 2015, superbolide over Romania and structural diversity of meter-sized asteroids}.
\newblock {\em Planetary and Space Science}, 143:147--158.

\bibitem[{Borovi{\v{c}}ka} et~al., 2020]{Borovicka2020AJ16042B}
{Borovi{\v{c}}ka}, J., {Spurn{\'y}}, P., and {Shrben{\'y}}, L. (2020).
\newblock {Two Strengths of Ordinary Chondritic Meteoroids as Derived from Their Atmospheric Fragmentation Modeling}.
\newblock {\em The Astronomical Journal}, 160(1):42.

\bibitem[{Borovi{\v{c}}ka} et~al., 2015b]{Borovicka2015MPS501244B}
{Borovi{\v{c}}ka}, J., {Spurn{\'y}}, P., {{\v{S}}egon}, D., {Andrei{\'c}}, {\v{Z}}., {Kac}, J., {Korlevi{\'c}}, K., {Atanackov}, J., {Kladnik}, G., {Mucke}, H., {Vida}, D., and {Novoselnik}, F. (2015b).
\newblock {The instrumentally recorded fall of the Kri{\v{z}}evci meteorite, Croatia, February 4, 2011}.
\newblock {\em Meteoritics \& Planetary Science}, 50(7):1244--1259.

\bibitem[{Borovi{\v{c}}ka} et~al., 2013]{Borovicka2013MPS481757B}
{Borovi{\v{c}}ka}, J., {T{\'o}th}, J., {Igaz}, A., {Spurn{\'y}}, P., {Kalenda}, P., {Haloda}, J., {Svore{\r{a}}}, J., {Korno{\v{s}}}, L., {Silber}, E., {Brown}, P., and {Hus{\'a}Rik}, M. (2013).
\newblock {The Ko{\v{s}}ice meteorite fall: Atmospheric trajectory, fragmentation, and orbit}.
\newblock {\em Meteoritics \& Planetary Science}, 48(10):1757--1779.

\bibitem[{Bro{\v{z}}} et~al., 2024]{Broz2024AA}
{Bro{\v{z}}}, M., {Vernazza}, P., {Marsset}, M., {Binzel}, R.~P., {DeMeo}, F., {Birlan}, M., {Colas}, F., {Anghel}, S., {Bouley}, S., {Blanpain}, C., {Gattacceca}, J., {Jeanne}, S., {Jorda}, L., {Lecubin}, J., {Malgoyre}, A., {Steinhausser}, A., {Vaubaillon}, J., and {Zanda}, B. (2024).
\newblock {Source regions of carbonaceous meteorites and near-Earth objects}.
\newblock {\em Astronomy \& Astrophysics}, 689:A183.

\bibitem[Brown et~al., 2024]{Brown2024}
Brown, M., Lewis, H., Kavanagh, A., Cnossen, I., and Elvidge, S. (2024).
\newblock Future climate change in the thermosphere under varying solar activity conditions.
\newblock {\em Journal of Geophysical Research: Space Physics}.

\bibitem[Brown et~al., 2021]{Brown2021}
Brown, M.~K., Lewis, H.~G., Kavanagh, A.~J., and Cnossen, I. (2021).
\newblock Future decreases in thermospheric neutral density in {L}ow {E}arth {O}rbit due to carbon dioxide emissions.
\newblock {\em Journal of Geophysical Research: Atmospheres}, 126(8).

\bibitem[Brown et~al., 2000]{Brown2000}
Brown, P.~G., Hildebrand, A.~R., Zolensky, M.~E., Grady, M., Clayton, R.~N., Mayeda, T.~K., Tagliaferri, E., Spalding, R., MacRae, N.~D., Hoffman, E.~L., Mittlefehldt, D.~W., Wacker, J.~F., Bird, J.~A., Campbell, M.~D., Carpenter, R., Gingerich, H., Glatiotis, M., Greiner, E., Mazur, M.~J., McCausland, P.~J., Plotkin, H., and Mazur, T.~R. (2000).
\newblock The fall, recovery, orbit, and composition of the tagish lake meteorite: A new type of carbonaceous chondrite.
\newblock {\em Science}, 290(5490):320--325.

\bibitem[{Brown} et~al., 2023]{Brown2023MPS581773B}
{Brown}, P.~G., {McCausland}, P.~J.~A., {Hildebrand}, A.~R., {Hanton}, L.~T.~J., {Eckart}, L.~M., {Busemann}, H., {Krietsch}, D., {Maden}, C., {Welten}, K., {Caffee}, M.~W., {Laubenstein}, M., {Vida}, D., {Ciceri}, F., {Silber}, E., {Herd}, C.~D.~K., {Hill}, P., {Devillepoix}, H., {Sansom}, E.~K., {Cup{\'a}k}, M., {Anderson}, S., {Flemming}, R.~L., {Nelson}, A.~J., {Mazur}, M., {Moser}, D.~E., {Cooke}, W.~J., {Hladiuk}, D., {Male{\v{c}}i{\'c}}, B., {Prtenjak}, M.~T., and {Nowell}, R. (2023).
\newblock {The Golden meteorite fall: Fireball trajectory, orbit, and meteorite characterization}.
\newblock {\em Meteoritics \& Planetary Science}, 58(12):1773--1807.

\bibitem[Brown et~al., 2002]{brown2002entry}
Brown, P.~G., Revelle, D.~O., Tagliaferri, E., and Hildebrand, A.~R. (2002).
\newblock An entry model for the tagish lake fireball using seismic, satellite and infrasound records.
\newblock {\em Meteoritics \& Planetary Science}, 37(5):661--675.

\bibitem[{Campbell-Burns} and {Kacerek}, 2014]{CampbellBurns2014JIMO42139C}
{Campbell-Burns}, P. and {Kacerek}, R. (2014).
\newblock {The UK Meteor Observation Network}.
\newblock {\em WGN, Journal of the International Meteor Organization}, 42(4):139--144.

\bibitem[{Ceplecha} et~al., 1998]{Ceplecha1998SSRv84327C}
{Ceplecha}, Z., {Borovi{\v{c}}ka}, J., {Elford}, W.~G., {Revelle}, D.~O., {Hawkes}, R.~L., {Porub{\v{c}}an}, V., and {{\v{S}}imek}, M. (1998).
\newblock {Meteor Phenomena and Bodies}.
\newblock {\em Space Science Reviews}, 84:327--471.

\bibitem[{Clemesha} and {Batista}, 2006]{Clemesha2006JASTP681934C}
{Clemesha}, B. and {Batista}, P. (2006).
\newblock {The quantification of long-term atmospheric change via meteor ablation height measurements}.
\newblock {\em Journal of Atmospheric and Solar-Terrestrial Physics}, 68(17):1934--1939.

\bibitem[Cnossen, 2020]{Cnossen2020}
Cnossen, I. (2020).
\newblock Analysis and attribution of climate change in the upper atmosphere from 1950 to 2015 simulated by {WACCM-X}.
\newblock {\em Journal of Geophysical Research: Space Physics}, 125(12).

\bibitem[Cnossen, 2022]{Cnossen2022}
Cnossen, I. (2022).
\newblock A realistic projection of climate change in the upper atmosphere into the 21st century.
\newblock {\em Geophysical Research Letters}, 49(19).

\bibitem[Cnossen et~al., 2024]{Cnossen2024}
Cnossen, I., Emmert, J.~T., Garcia, R.~R., Elias, A.~G., Mlynczak, M.~G., and Zhang, S.-R. (2024).
\newblock A review of global long-term changes in the mesosphere, thermosphere and ionosphere: A starting point for inclusion in (semi-) empirical models.
\newblock {\em Advances in Space Research}, 74(11):5991--6011.

\bibitem[{Colas} et~al., 2020]{Colas2020AA644A53C}
{Colas}, F., {Zanda}, B., {Bouley}, S., {Jeanne}, S., {Malgoyre}, A., {Birlan}, M., {Blanpain}, C., {Gattacceca}, J., {Jorda}, L., {Lecubin}, J., {Marmo}, C., {Rault}, J.~L., {Vaubaillon}, J., {Vernazza}, P., {Yohia}, C., {Gardiol}, D., {Nedelcu}, A., {Poppe}, B., {Rowe}, J., {Forcier}, M., {Koschny}, D., {Trigo-Rodriguez}, J.~M., {Lamy}, H., {Behrend}, R., {Ferri{\`e}re}, L., {Barghini}, D., {Buzzoni}, A., {Carbognani}, A., {Di Carlo}, M., {Di Martino}, M., {Knapic}, C., {Londero}, E., {Pratesi}, G., {Rasetti}, S., {Riva}, W., {Stirpe}, G.~M., {Valsecchi}, G.~B., {Volpicelli}, C.~A., {Zorba}, S., {Coward}, D., {Drolshagen}, E., {Drolshagen}, G., {Hernandez}, O., {Jehin}, E., {Jobin}, M., {King}, A., {Nitschelm}, C., {Ott}, T., {Sanchez-Lavega}, A., {Toni}, A., {Abraham}, P., {Affaticati}, F., {Albani}, M., {Andreis}, A., {Andrieu}, T., {Anghel}, S., {Antaluca}, E., {Antier}, K., {App{\'e}r{\'e}}, T., {Armand}, A., {Ascione}, G., {Audureau}, Y., {Auxepaules}, G., {Avoscan}, T., {Baba Aissa}, D., {Bacci}, P.,
  {B{\v{a}}descu}, O., {Baldini}, R., {Baldo}, R., {Balestrero}, A., {Baratoux}, D., {Barbotin}, E., {Bardy}, M., {Basso}, S., {Bautista}, O., {Bayle}, L.~D., {Beck}, P., {Bellitto}, R., {Belluso}, R., {Benna}, C., {Benammi}, M., {Beneteau}, E., {Benkhaldoun}, Z., {Bergamini}, P., {Bernardi}, F., {Bertaina}, M.~E., {Bessin}, P., {Betti}, L., {Bettonvil}, F., {Bihel}, D., {Birnbaum}, C., {Blagoi}, O., {Blouri}, E., {Boac{\u{a}}}, I., {Boat{\v{a}}}, R., {Bobiet}, B., {Bonino}, R., {Boros}, K., {Bouchet}, E., {Borgeot}, V., {Bouchez}, E., {Boust}, D., {Boudon}, V., {Bouman}, T., {Bourget}, P., {Brandenburg}, S., {Bramond}, P., {Braun}, E., {Bussi}, A., {Cacault}, P., {Caillier}, B., {Calegaro}, A., {Camargo}, J., {Caminade}, S., {Campana}, A.~P.~C., {Campbell-Burns}, P., {Canal-Domingo}, R., {Carell}, O., {Carreau}, S., {Cascone}, E., {Cattaneo}, C., {Cauhape}, P., {Cavier}, P., {Celestin}, S., {Cellino}, A., {Champenois}, M., {Chennaoui Aoudjehane}, H., {Chevrier}, S., {Cholvy}, P., {Chomier}, L., {Christou},
  A., {Cricchio}, D., {Coadou}, P., {Cocaign}, J.~Y., {Cochard}, F., {Cointin}, S., {Colombi}, E., {Colque Saavedra}, J.~P., {Corp}, L., {Costa}, M., {Costard}, F., {Cottier}, M., {Cournoyer}, P., {Coustal}, E., {Cremonese}, G., {Cristea}, O., {Cuzon}, J.~C., {D'Agostino}, G., {Daiffallah}, K., {D{\v{a}}nescu}, C., {Dardon}, A., {Dasse}, T., {Davadan}, C., {Debs}, V., {Defaix}, J.~P., {Deleflie}, F., {D'Elia}, M., {De Luca}, P., {De Maria}, P., {Deverch{\`e}re}, P., {Devillepoix}, H., {Dias}, A., {Di Dato}, A., {Di Luca}, R., {Dominici}, F.~M., {Drouard}, A., {Dumont}, J.~L., {Dupouy}, P., {Duvignac}, L., {Egal}, A., {Erasmus}, N., {Esseiva}, N., {Ebel}, A., {Eisengarten}, B., {Federici}, F., {Feral}, S., {Ferrant}, G., {Ferreol}, E., {Finitzer}, P., {Foucault}, A., {Francois}, P., {Fr{\^\i}ncu}, M., {Froger}, J.~L., {Gaborit}, F., {Gagliarducci}, V., {Galard}, J., {Gardavot}, A., {Garmier}, M., {Garnung}, M., {Gautier}, B., {Gendre}, B., {Gerard}, D., {Gerardi}, A., {Godet}, J.~P., {Grandchamps}, A.,
  {Grouiez}, B., {Groult}, S., {Guidetti}, D., {Giuli}, G., and {Hello}, Y. (2020).
\newblock {FRIPON: a worldwide network to track incoming meteoroids}.
\newblock {\em Astronomy \& Astrophysics}, 644:A53.

\bibitem[{Daly} et~al., 2020]{Daly2020EPSC14705D}
{Daly}, L., {McMullan}, S., {Rowe}, J., {Collins}, G.~S., {Suttle}, M., {Chan}, Q. H.~S., {Young}, J.~S., {Shaw}, C., {Mardon}, A.~G., {Alexander}, M., {Tate}, J., {The Desert Fireball Network Team}, {Campbell-Burns}, P., {Kacerek}, R., {King}, A., {Joy}, K., {Christou}, A., {Hor{\'a}k}, J., and {Shepherd}, J. (2020).
\newblock {The UK Fireball Alliance (UKFAll); combining and integrating the diversity of UK camera networks to aim to recover the first UK meteorite fall for 30 years}.
\newblock In {\em European Planetary Science Congress}, pages EPSC2020--705.

\bibitem[Dawkins et~al., 2023]{Dawkins2023}
Dawkins, E. C.~M., Stober, G., Janches, D., Carrillo‐Sánchez, J.~D., Lieberman, R.~S., Jacobi, C., Moffat‐Griffin, T., Mitchell, N.~J., Cobbett, N., Batista, P.~P., Andrioli, V.~F., Buriti, R.~A., Murphy, D.~J., Kero, J., Gulbrandsen, N., Tsutsumi, M., Kozlovsky, A., Kim, J.~H., Lee, C., and Lester, M. (2023).
\newblock Solar cycle and long‐term trends in the observed peak of the meteor altitude distributions by meteor radars.
\newblock {\em Geophysical Research Letters}, 50(2).

\bibitem[{Devillepoix} et~al., 2020]{Devillepoix2020PSS19105036D}
{Devillepoix}, H.~A.~R., {Cup{\'a}k}, M., {Bland}, P.~A., {Sansom}, E.~K., {Towner}, M.~C., {Howie}, R.~M., {Hartig}, B.~A.~D., {Jansen-Sturgeon}, T., {Shober}, P.~M., {Anderson}, S.~L., {Benedix}, G.~K., {Busan}, D., {Sayers}, R., {Jenniskens}, P., {Albers}, J., {Herd}, C.~D.~K., {Hill}, P.~J.~A., {Brown}, P.~G., {Krzeminski}, Z., {Osinski}, G.~R., {Aoudjehane}, H.~C., {Benkhaldoun}, Z., {Jabiri}, A., {Guennoun}, M., {Barka}, A., {Darhmaoui}, H., {Daly}, L., {Collins}, G.~S., {McMullan}, S., {Suttle}, M.~D., {Ireland}, T., {Bonning}, G., {Baeza}, L., {Alrefay}, T.~Y., {Horner}, J., {Swindle}, T.~D., {Hergenrother}, C.~W., {Fries}, M.~D., {Tomkins}, A., {Langendam}, A., {Rushmer}, T., {O'Neill}, C., {Janches}, D., {Hormaechea}, J.~L., {Shaw}, C., {Young}, J.~S., {Alexander}, M., {Mardon}, A.~D., and {Tate}, J.~R. (2020).
\newblock {A Global Fireball Observatory}.
\newblock {\em Planetary and Space Science}, 191:105036.

\bibitem[Emmert, 2015]{Emmert2015}
Emmert, J.~T. (2015).
\newblock Altitude and solar activity dependence of 1967-2005 thermospheric density trends derived from orbital drag.
\newblock {\em Journal of Geophysical Research: Space Physics}, 120(4):2940--2950.

\bibitem[{Gattacceca} et~al., 2022]{Gattacceca2022MPS572102G}
{Gattacceca}, J., {MCCubbin}, F.~M., {Grossman}, J., {Bouvier}, A., {Chabot}, N.~L., {D'Orazio}, M., {Goodrich}, C., {Greshake}, A., {Gross}, J., {Komatsu}, M., {Miao}, B., and {Schrader}, D. (2022).
\newblock {The Meteoritical Bulletin, No. 110}.
\newblock {\em Meteoritics \& Planetary Science}, 57(11):2102--2105.

\bibitem[{Hankey} et~al., 2020]{Hankey2020pimoconf21H}
{Hankey}, M., {Meisel}, D., and {Perlerin}, V. (2020).
\newblock {The All-Sky-6 and Video Meteor Archive System of the AMS Ltd.}
\newblock In {Pajer}, U., {Rendtel}, J., {Gyssens}, M., and {Verbeeck}, C., editors, {\em International Meteor Conference, Bollmannsruh, Germany}, pages 21--25.

\bibitem[{Hromakina} et~al., 2021]{Hromakina2021AA656A89H}
{Hromakina}, T., {Birlan}, M., {Barucci}, M.~A., {Fulchignoni}, M., {Colas}, F., {Fornasier}, S., {Merlin}, F., {Sonka}, A., {Petrescu}, E., {Perna}, D., {Dotto}, E., and {Neorocks T Eam} (2021).
\newblock {Photometric survey of 55 near-earth asteroids}.
\newblock {\em Astronomy \& Astrophysics}, 656:A89.

\bibitem[{Ieva} et~al., 2020]{Ieva2020AA644A23I}
{Ieva}, S., {Dotto}, E., {Mazzotta Epifani}, E., {Perna}, D., {Fanasca}, C., {Lazzarin}, M., {Bertini}, I., {Petropoulou}, V., {Rossi}, A., {Micheli}, M., and {Perozzi}, E. (2020).
\newblock {Extended photometric survey of near-Earth objects}.
\newblock {\em Astronomy \& Astrophysics}, 644:A23.

\bibitem[{Jacobi}, 2014]{Jacobi2014AdRS12161J}
{Jacobi}, C. (2014).
\newblock {Meteor heights during the recent solar minimum}.
\newblock {\em Advances in Radio Science}, 12:161--165.

\bibitem[{King} et~al., 2022]{King2022SciA8Q3925K}
{King}, A.~J., {Daly}, L., {Rowe}, J., {Joy}, K.~H., {Greenwood}, R.~C., {Devillepoix}, H. A.~R., {Suttle}, M.~D., {Chan}, Q. H.~S., {Russell}, S.~S., {Bates}, H.~C., {Bryson}, J. F.~J., {Clay}, P.~L., {Vida}, D., {Lee}, M.~R., {O'Brien}, {\'A}., {Hallis}, L.~J., {Stephen}, N.~R., {Tart{\`e}se}, R., {Sansom}, E.~K., {Towner}, M.~C., {Cupak}, M., {Shober}, P.~M., {Bland}, P.~A., {Findlay}, R., {Franchi}, I.~A., {Verchovsky}, A.~B., {Abernethy}, F. A.~J., {Grady}, M.~M., {Floyd}, C.~J., {Van Ginneken}, M., {Bridges}, J., {Hicks}, L.~J., {Jones}, R.~H., {Mitchell}, J.~T., {Genge}, M.~J., {Jenkins}, L., {Martin}, P.-E., {Sephton}, M.~A., {Watson}, J.~S., {Salge}, T., {Shirley}, K.~A., {Curtis}, R.~J., {Warren}, T.~J., {Bowles}, N.~E., {Stuart}, F.~M., {Di Nicola}, L., {Gy{\"o}re}, D., {Boyce}, A.~J., {Shaw}, K. M.~M., {Elliott}, T., {Steele}, R. C.~J., {Povinec}, P., {Laubenstein}, M., {Sanderson}, D., {Cresswell}, A., {Jull}, A. J.~T., {S{\'y}kora}, I., {Sridhar}, S., {Harrison}, R.~J., {Willcocks}, F.~M.,
  {Harrison}, C.~S., {Hallatt}, D., {Wozniakiewicz}, P.~J., {Burchell}, M.~J., {Alesbrook}, L.~S., {Dignam}, A., {Almeida}, N.~V., {Smith}, C.~L., {Clark}, B., {Humphreys-Williams}, E.~R., {Schofield}, P.~F., {Cornwell}, L.~T., {Spathis}, V., {Morgan}, G.~H., {Perkins}, M.~J., {Kacerek}, R., {Campbell-Burns}, P., {Colas}, F., {Zanda}, B., {Vernazza}, P., {Bouley}, S., {Jeanne}, S., {Hankey}, M., {Collins}, G.~S., {Young}, J.~S., {Shaw}, C., {Horak}, J., {Jones}, D., {James}, N., {Bosley}, S., {Shuttleworth}, A., {Dickinson}, P., {McMullan}, I., {Robson}, D., {Smedley}, A. R.~D., {Stanley}, B., {Bassom}, R., {McIntyre}, M., {Suttle}, A.~A., {Fleet}, R., {Bastiaens}, L., {Ih{\'a}sz}, M.~B., {McMullan}, S., {Boazman}, S.~J., {Dickeson}, Z.~I., {Grindrod}, P.~M., {Pickersgill}, A.~E., {Weir}, C.~J., {Suttle}, F.~M., {Farrelly}, S., {Spencer}, I., {Naqvi}, S., {Mayne}, B., {Skilton}, D., {Kirk}, D., {Mounsey}, A., {Mounsey}, S.~E., {Mounsey}, S., {Godfrey}, P., {Bond}, L., {Bond}, V., {Wilcock}, C., {Wilcock}, H.,
  and {Wilcock}, R. (2022).
\newblock {The Winchcombe meteorite, a unique and pristine witness from the outer solar system}.
\newblock {\em Science Advances}, 8(46):eabq3925.

\bibitem[{Koschny} and {Borovicka}, 2017]{Koschny2017JIMO4591K}
{Koschny}, D. and {Borovicka}, J. (2017).
\newblock {Definitions of terms in meteor astronomy}.
\newblock {\em WGN, Journal of the International Meteor Organization}, 45(5):91--92.

\bibitem[{Lima} et~al., 2015]{Lima2015JASTP133139L}
{Lima}, L.~M., {Ara{\'u}jo}, L.~R., {Alves}, E.~O., {Batista}, P.~P., and {Clemesha}, B.~R. (2015).
\newblock {Variations in meteor heights at 22.7{\textdegree}S during solar cycle 23}.
\newblock {\em Journal of Atmospheric and Solar-Terrestrial Physics}, 133:139--144.

\bibitem[Lyytinen and Gritsevich, 2016]{LyytinenGritsevich2016}
Lyytinen, E. and Gritsevich, M. (2016).
\newblock Implications of the atmospheric density profile in the processing of fireball observations.
\newblock {\em Planetary and Space Science}, 120:35--42.

\bibitem[Manabe and Wetherald, 1967]{Manabe1967}
Manabe, S. and Wetherald, R.~T. (1967).
\newblock Thermal equilibrium of the atmosphere with a given distribution of relative humidity.
\newblock {\em Journal of the Atmospheric Sciences}, 24(3):241--259.

\bibitem[Manabe and Wetherald, 1975]{Manabe1975}
Manabe, S. and Wetherald, R.~T. (1975).
\newblock The effects of doubling the {CO\textsubscript{2}} concentration on the climate of a general circulation model.
\newblock {\em Journal of the Atmospheric Sciences}, 32(1):3--15.

\bibitem[Matthes et~al., 2017]{Matthes2017}
Matthes, K., Funke, B., Andersson, M.~E., Barnard, L., Beer, J., Charbonneau, P., Clilverd, M.~A., de~Wit, T.~D., Haberreiter, M., Hendry, A., Jackman, C.~H., Kretzschmar, M., Kruschke, T., Kunze, M., Langematz, U., Marsh, D.~R., Maycock, A.~C., Misios, S., Rodger, C.~J., Scaife, A.~A., Seppälä, A., Shangguan, M., Sinnhuber, M., Tourpali, K., Usoskin, I., van~de Kamp, M., Verronen, P.~T., and Versick, S. (2017).
\newblock Solar forcing for {CMIP}6 (v3.2).
\newblock {\em Geoscientific Model Development}, 10(6):2247--2302.

\bibitem[{McMullan} et~al., 2024]{McMullan2024MPS59927M}
{McMullan}, S., {Vida}, D., {Devillepoix}, H. A.~R., {Rowe}, J., {Daly}, L., {King}, A.~J., {Cup{\'a}k}, M., {Howie}, R.~M., {Sansom}, E.~K., {Shober}, P., {Towner}, M.~C., {Anderson}, S., {McFadden}, L., {Hor{\'a}k}, J., {Smedley}, A. R.~D., {Joy}, K.~H., {Shuttleworth}, A., {Colas}, F., {Zanda}, B., {O'Brien}, {\'A}.~C., {McMullan}, I., {Shaw}, C., {Suttle}, A., {Suttle}, M.~D., {Young}, J.~S., {Campbell-Burns}, P., {Kacerek}, R., {Bassom}, R., {Bosley}, S., {Fleet}, R., {Jones}, D., {McIntyre}, M., {James}, N., {Robson}, D., {Dickinson}, P., {Bland}, P.~A., and {Collins}, G.~S. (2024).
\newblock {The Winchcombe fireball{\textemdash}That lucky survivor}.
\newblock {\em Meteoritics and Planetary Science}, 59(5):927--947.

\bibitem[O'Neill et~al., 2016]{ONeill2016}
O'Neill, B.~C., Tebaldi, C., van Vuuren, D.~P., Eyring, V., Friedlingstein, P., Hurtt, G., Knutti, R., Kriegler, E., Lamarque, J.-F., Lowe, J., Meehl, G.~A., Moss, R., Riahi, K., and Sanderson, B.~M. (2016).
\newblock The scenario model intercomparison project ({ScenarioMIP}) for {CMIP}6.
\newblock {\em Geoscientific Model Development}, 9(9):3461--3482.

\bibitem[{Picone} et~al., 2002]{Picone2002JGRA1071468P}
{Picone}, J.~M., {Hedin}, A.~E., {Drob}, D.~P., and {Aikin}, A.~C. (2002).
\newblock {NRLMSISE-00 empirical model of the atmosphere: Statistical comparisons and scientific issues}.
\newblock {\em Journal of Geophysical Research (Space Physics)}, 107(A12):1468.

\bibitem[Roble and Dickinson, 1989]{Roble1989}
Roble, R.~G. and Dickinson, R.~E. (1989).
\newblock How will changes in carbon dioxide and methane modify the mean structure of the mesosphere and thermosphere?
\newblock {\em Geophysical Research Letters}, 16(12):1441--1444.

\bibitem[{Russell} et~al., 2024]{Russell2024MPS59973R}
{Russell}, S.~S., {King}, A.~J., {Bates}, H.~C., {Almeida}, N.~V., {Greenwood}, R.~C., {Daly}, L., {Joy}, K.~H., {Rowe}, J., {Salge}, T., {Smith}, C.~L., {Grindrod}, P., {Boazman}, S., {Bond}, L., {Bond}, V., {Casey}, C., {Dickeson}, Z., {Ensor}, G., {Farrelly}, S., {Godfrey}, P., {Hallis}, L.~J., {Ih{\'a}sz}, M.~B., {Kirk}, D., {Jackson}, L., {Lee}, M.~R., {Mayne}, B., {MCMullan}, S., {Mounsey}, A., {Mounsey}, S.~E., {Mounsey}, S., {Motaghian}, S., {Naqvi}, S., {O'Brien}, {\'A}., {Pickersgill}, A., {Skilton}, D., {Spencer}, I., {Stephen}, N.~R., {Suttle}, F., {Suttle}, M.~D., {Tartese}, R., {Weir}, C., {Wilcock}, C., {Wilcock}, H., and {Wilcock}, R. (2024).
\newblock {Recovery and curation of the Winchcombe (CM2) meteorite}.
\newblock {\em Meteoritics \& Planetary Science}, 59(5):973--987.

\bibitem[Shober et~al., 2025]{Shober2025}
Shober, P.~M., Devillepoix, H. A.~R., Vaubaillon, J., Anghel, S., Deam, S.~E., Sansom, E.~K., Colas, F., Zanda, B., Vernazza, P., and Bland, P. (2025).
\newblock Perihelion history and atmospheric survival as primary drivers of the earth's meteorite record.
\newblock {\em Nature Astronomy}, 9(6):799--812.

\bibitem[{Stewart} et~al., 2013]{Stewart2013JIMO4184S}
{Stewart}, W., {Pratt}, A.~R., and {Entwisle}, L. (2013).
\newblock {NEMETODE: The Network for Meteor Triangulation and Orbit Determination. System Overview and Initial Results from a UK Video Meteor Network}.
\newblock {\em WGN, Journal of the International Meteor Organization}, 41(3):84--91.

\bibitem[{Stober} et~al., 2012]{Stober2012JASTP7455S}
{Stober}, G., {Jacobi}, C., {Matthias}, V., {Hoffmann}, P., and {Gerding}, M. (2012).
\newblock {Neutral air density variations during strong planetary wave activity in the mesopause region derived from meteor radar observations}.
\newblock {\em Journal of Atmospheric and Solar-Terrestrial Physics}, 74:55--63.

\bibitem[{Stober} et~al., 2014]{Stober2014GeoRL416919S}
{Stober}, G., {Matthias}, V., {Brown}, P., and {Chau}, J.~L. (2014).
\newblock {Neutral density variation from specular meteor echo observations spanning one solar cycle}.
\newblock {\em Geophysical Research Letters}, 41(19):6919--6925.

\bibitem[{Suttle} et~al., 2024]{Suttle2024MPS591043S}
{Suttle}, M.~D., {Daly}, L., {Jones}, R.~H., {Jenkins}, L., {van Ginneken}, M., {Mitchell}, J.~T., {Bridges}, J.~C., {Hicks}, L.~J., {Johnson}, D., {Rollinson}, G., {Taylor}, R., {Genge}, M.~J., {Schr{\"o}der}, C., {Trimby}, P., {Mansour}, H., {Piazolo}, S., {Bonsall}, E., {Salge}, T., {Heard}, R., {Findlay}, R., {King}, A.~J., {Bates}, H.~C., {Lee}, M.~R., {Stephen}, N.~R., {Willcocks}, F.~M., {Greenwood}, R.~C., {Franchi}, I.~A., {Russell}, S.~S., {Harrison}, C.~S., {Schofield}, P.~F., {Almeida}, N.~V., {Floyd}, C., {Martin}, P.~E., {Joy}, K.~H., {Wozniakiewicz}, P.~J., {Hallatt}, D., {Burchell}, M.~J., {Alesbrook}, L.~S., {Spathis}, V., {Cornwell}, L.~T., and {Dignam}, A. (2024).
\newblock {The Winchcombe meteorite{\textemdash}A regolith breccia from a rubble pile CM chondrite asteroid}.
\newblock {\em Meteoritics \& Planetary Science}, 59(5):1043--1067.

\bibitem[{Venkat Ratnam} et~al., 2024]{VenkatRatnam2024}
{Venkat Ratnam}, M., Teja, A., Pramitha, M., Eswaraiah, S., and {Vijaya Bhaskara Rao}, S. (2024).
\newblock Climatology of meteor echoes and mean winds in the mlt region revealed by svu meteor radar over tirupati (13.63on, 79.4oe): Long-term trends.
\newblock {\em Advances in Space Research}.

\bibitem[Vida et~al., 2021]{vida2021high}
Vida, D., Brown, P.~G., Campbell-Brown, M., Weryk, R.~J., Stober, G., and McCormack, J.~P. (2021).
\newblock High precision meteor observations with the canadian automated meteor observatory: Data reduction pipeline and application to meteoroid mechanical strength measurements.
\newblock {\em Icarus}, 354:114097.

\bibitem[{Vida} et~al., 2023]{Vida2023NatAs7318V}
{Vida}, D., {Brown}, P.~G., {Devillepoix}, H. A.~R., {Wiegert}, P., {Moser}, D.~E., {Matlovi{\v{c}}}, P., {Herd}, C. D.~K., {Hill}, P. J.~A., {Sansom}, E.~K., {Towner}, M.~C., {T{\'o}th}, J., {Cooke}, W.~J., and {Hladiuk}, D.~W. (2023).
\newblock {Direct measurement of decimetre-sized rocky material in the Oort cloud}.
\newblock {\em Nature Astronomy}, 7:318--329.

\bibitem[{Vida} et~al., 2021]{Vida2021MNRAS5065046V}
{Vida}, D., {{\v{S}}egon}, D., {Gural}, P.~S., {Brown}, P.~G., {McIntyre}, M. J.~M., {Dijkema}, T.~J., {Pavleti{\'c}}, L., {Kuki{\'c}}, P., {Mazur}, M.~J., {Eschman}, P., {Roggemans}, P., {Merlak}, A., and {Zubovi{\'c}}, D. (2021).
\newblock {The Global Meteor Network - Methodology and first results}.
\newblock {\em MNRAS}, 506(4):5046--5074.

\bibitem[Weng et~al., 2020]{Weng2020}
Weng, L., Lei, J., Zhong, J., Dou, X., and Fang, H. (2020).
\newblock A machine-learning approach to derive long-term trends of thermospheric density.
\newblock {\em Geophysical Research Letters}, 47(6).

\end{thebibliography}

\end{document}